\documentclass[hidelinks,onefignum,onetabnum, notocite]{siamart220329}

\usepackage{lipsum}
\usepackage{amsfonts}
\usepackage{graphicx}
\usepackage{epstopdf}
\usepackage{algorithmic}
\usepackage{physics}
\usepackage{array}
\usepackage{enumitem}

\ifpdf
  \DeclareGraphicsExtensions{.eps,.pdf,.png,.jpg}
\else
  \DeclareGraphicsExtensions{.eps}
\fi

\newsiamremark{remark}{Remark}
\newsiamremark{hypothesis}{Hypothesis}
\crefname{hypothesis}{Hypothesis}{Hypotheses}
\newsiamthm{claim}{Claim}

\headers{Deterministic Fast and Stable Phase Retrieval in Multiple Dimensions}{C. Brabec, S. Trajtenberg-Mills, L. Daniel, D. Englund}

\title{Deterministic Fast and Stable Phase Retrieval in Multiple Dimensions\thanks{Submitted to the editors DATE.
\funding{This material is based upon work supported by the National Science Foundation Graduate Research Fellowship Program under Grant No. 2141064. STM is supported by the MITRE Quantum Moonshot Program.}}}

\author{Cole Brabec\thanks{RLE, MIT
  (\email{cbrabec@mit.edu}).}
\and Sivan Trajtenberg-Mills
\and Luca Daniel 
\and Dirk Englund }

\usepackage{amsopn}

\ifpdf
\hypersetup{
  pdftitle={Fast and Stable Phase Retrieval in Arbitrary Dimensions via the Schwarz Transform},
  pdfauthor={C. Brabec, S. Trajtenberg-Mills, D. Englund, L. Daniel}
}
\fi

\begin{document}

\maketitle

\begin{abstract}
We present the first phase retrieval algorithm guaranteed to solve the multidimensional phase retrieval problem in polynomial arithmetic complexity without prior information. The method succesfully terminates in $\Tilde{O}(N \log(N))$ operations for Fourier measurements with cardinality N. The algorithm is guaranteed to succeed for a large class of objects, which we term "Schwarz objects". We further present an easy-to-calculate and well-conditioned diagonal operator that transforms any feasible phase-retrieval instance into one that is solved by our method. We derive our method by combining techniques from classical complex analysis, algebraic topology, and modern numerical analysis. Concretely, we pose the phase retrieval problem as a multiplicative Cousin problem, construct an approximate solution using a modified integral used for the Schwarz problem, and refine the approximate solution to an exact solution via standard optimization methods. We present numerical experimentation demonstrating our algorithm’s performance and its superiority to existing method. Finally, we demonstrate that our method is robust against Gaussian noise.

\end{abstract}



\section{Introduction}
\subsection{Problem Introduction}
\begin{figure}[H]
    \centering
\includegraphics[scale = .45]{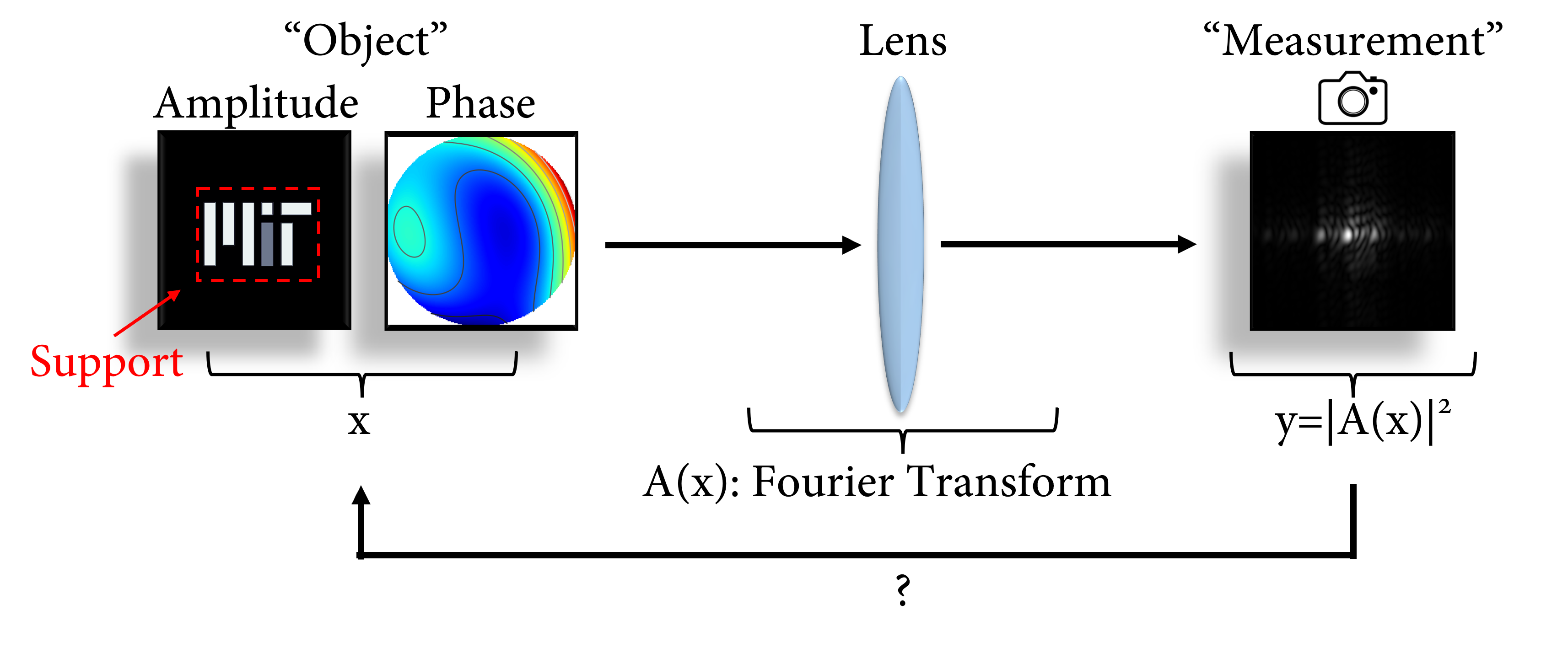}
    \caption{An example phase retrieval experiment.}
    \label{fig:schematic}
\end{figure}

A century-old problem plaguing physicists, engineers, and applied mathematicians alike has been the recovery of a complex vector from the magnitude of its Fourier transform. The general form of this problem, known as \textit{Phase Retrieval}, has proven crucial in applications ranging from medical imaging and crystallography to quantum state tomography \cite{shechtman_phase_2015, jaganathan_phase_2016, bendory_toward_2020-1}. In the canonical phase retrieval setting, an electromagnetic field with an unknown phase and magnitude distribution undergoes a Fourier transform by propagating through free space or a lens and is then measured by a camera (\cref{fig:schematic}). Since a camera captures only intensity information, the phase retrieval problem must be solved to recover the original phase and magnitude profile.

Generalizing from the Fourier transform to a general linear operator $\mathcal{A}$, we formally pose phase retrieval as the following quadratic feasibility problem:
\begin{equation}\label{PhaseRet}
    \text{find}_{\textbf{x} \in \mathcal{C}} \quad |\mathcal{A} \{ \textbf{x}\}|^2 = \textbf{y} \tag{PR}
\end{equation}

To provide robustness against noise, \cref{PhaseRet} is normally solved by minimizing an appropriate penalty function. In this work, we focus on the least-squares penalty:
\begin{equation}\label{eq:LSRet}
    \min_{\textbf{x} \in \mathcal{C}} \quad \norm{|\mathcal{A}\{\textbf{x}\}|^2 - \textbf{y}}_2^2 
\end{equation}
We consider the discrete version of the problem in d dimensions. In this setting, the object to be recovered $\textbf{x}$ is a multidimensional complex vector  $\textbf{x} \in \mathbb{C}^{\textbf{n}} = \mathbb{C}^{n_1 \times n_2 \times ... \times n_d}$, $\mathcal{A} \in \mathcal{L}(\mathbb{C}^\textbf{n}, \mathbb{C}^{\textbf{m}})$, $\mathcal{C} \subset \mathbb{C}^{\textbf{n}}$, $\textbf{y} \in \mathbb{R}_{> 0}^{\textbf{m}}$, $\textbf{n}, \textbf{m} \in \mathbb{N}^d$ (We make frequent use of multi-index notation; unfamiliar readers can refer to \cref{sec: Notation}). 
Given the problem's physical origin, $\textbf{y}$ is classically referred to as the "measurement" and $\textbf{x}$ is referred to as the "object". 
The linear operator $\mathcal{A}$ represents the physical transformation induced by a given experimental setup, and the constraint set $\mathcal{C}$ represents some prior knowledge of the object to be recovered.

 The choice of operator $\mathcal{A}$ is informed by a mix of practical and mathematical considerations. For imaging applications, $\mathcal{A}$ is almost universally given by a direct sum of the form $ \bigoplus_i \mathcal{F} \mathcal{D}_i$, where the $\mathcal{D}_i$ are diagonal operators representing structured illumination. Structured illumination is used both to guarantee the existence of a unique solution to \cref{PhaseRet}, and to make the accompanying minimization problem tractable. Uniqueness is most often required only up to the fundamental symmetries of the Fourier operator: circular shifts, scalar multiplication by unimodular constant, and conjugate reflections. Some of the most popular choices of illumination are random and Fresnel phase masks. Randomized phase masks in particular have been shown to greatly speed numerical convergence \cite{candes_phase_2015}. However, these masks are often challenging to synthesize in experimental setting, and characterizing such an illumination often requires solving a different instance of phase retrieval. From a physical practicality perspective it is desirable to keep $\mathcal{A}$ as close to $\mathcal{F}$ as possible.
An additional degree of freedom is the oversampling ratio, or sample complexity, $M/N$. Here again the desires of algorithm designers and experimentalists differ. Most methods require a large $M/N$, sometimes in excess of 10 or more, to guarantee recovery \cite{shechtman_phase_2015}. However, small oversampling ratios are sometimes necessitated by a given experimental apparatus. 

The remaining choice is that of the constraint set $\mathcal{C}$. There are two predominant types: sparsity constraints --- which limit the number of nonzero elements of an object --- and ``support'' constraints - a constraint on the size of the support of the object. Support constraints dominate the imaging community, being one of the original constraints considered for the problem \cite{fienup_phase_1982}. Sparsity constraints are primarily used in x-ray crystallography because of the low density of atoms \cite{bendory_toward_2020-1}, and will not be considered in this work.

\subsection{Performance Criteria and Prior Art}
Three criteria are commonly used in the literature to measure the performance of phase retrieval algorithms:
\begin{description}
\item[C1: Sample Complexity] The required number of of measurements $M$ to recover a an object, defined as a function of $N$. The optimal bound is $O(N)$ \cite{shechtman_phase_2015}.
\item[C2: Arithmetic Complexity] The required number of arithmetic operations to return a solution. For $\mathcal{A} = \mathcal{F}$, the optimal bound is $\Tilde{O}(N \log(N))$ and represents the cost of applying a DFT. 
\item[C3: Deterministic Success] Whether an algorithm deterministically returns the correct solution. For $\mathcal{A} = \mathcal{F}$, there should exist a well-defined class of \cref{PhaseRet} instances for which the algorithm is guaranteed to return the unique solution.
\end{description}
 Meeting the optimal bounds for these criteria is essential to achieve practical performance on very large problems. Even an extra poly-logarithmic factor is a serious penalty when dealing with millions of variables. Criteria C3 measures the ultimate reliability of the algorithm. Without a deterministic guarantee, a method cannot be used for accuracy critical applications. 

Table \ref{tab:prevApproaches} summarizes the performance of several classes of current state-of-the-art algorithms.  which we describe below

\begin{table}[]
    \centering
    \begin{tabular}{ |m{3.3cm}||m{2.3cm}|m{3cm}|m{2.2cm}| }
    \hline
         Method & Sample Complexity & Arithmetic Complexity & Deterministic Success   \\
         \hline
         CAD \cite{arnon_cylindrical_1984} & $O(N)$ & $\Tilde{O}(\exp(\exp(N)))$ & Yes  \\ [5 pt]
         \hline
         Matrix Lifting \cite{candes_phase_2015,candes_solving_2014,candes_phaselift_2013, goldstein_phasemax_2018} & $O(N \log(N))$ & $\Tilde{O}(N^3 \log(N))$ & No\\
         \hline
         Wirtinger Flow \cite{cai_optimal_2016, candes_phase_2015,wang_solving_2018,yonel_generalization_2019} & $O(N \log(N))$ & $\Tilde{O}(N \log(N))$ & No \\
         \hline
         This Work & $O(N)$ & $\Tilde{O}(N \log(N))$ & Yes\\
         \hline
    \end{tabular}
    \caption{Summary of Prior Art. All values are specified for $\mathcal{A} = \mathcal{F}$.}
    \label{tab:prevApproaches}
\end{table}

First, Cylindrical Algebraic Decomposition (CAD) is an implementation of \hfill \break Tarski's quantifier elimination, which will return all solutions for any set of polynomial equations, such as those defining \cref{PhaseRet} \cite{arnon_cylindrical_1984}. However, the doubly exponential running time of CAD renders this algorithm completely impractical for all but the smallest problem instances. 

The next class of algorithms are matrix lifting methods \cite{candes_phase_2015}, first pioneered in PhaseLift. Underappreciated in the literature is the fact that given enough time, these algorithms in theory should be able to find a solution to any uniquely solvable \cref{PhaseRet} instance via the Laserre hierarchy \cite{josz_lasserre_2018}. However, the amount of resources required by matrix lifting grows very large in terms of size, required number of measurements, and run-time. Thus, matrix methods are impractical on medium-to-large problem instances. A further complication is that the approach generally does not work for $\mathcal{A} = \mathcal{F}$. Other measurement operators, such as random Gaussian operators or direct measurements of $|\textbf{x}|$ are needed for matrix lifting methods to succeed. 

Finally, a large number of methods perform some type of gradient descent on \cref{eq:LSRet}.  Wirtinger Flow \cite{candes_phase_2015} is the most widely known example of these methods and derivatives of Wirtinger Flow, such as Truncated Amplitude Flow \cite{wang_solving_2018}, are considered state-of-the-art. Existing gradient methods, like matrix methods, gradient methods lack a deterministic guarantee. The initialization methods used in the literature offer only probabilistic guarantees and rely on Gaussian non-Fourier measurements. The correctness proofs for linear sample complexity fundamentally break down for the case $\mathcal{A} = \mathcal{F}$. For Fourier measurements, the best known sample complexity is $O(N \log(N))$ for existing methods and require the use of coded diffraction patterns \cite{candes_phase_2015-4}. 

In this work, we introduce the first polynomial-time deterministic method to solve $\cref{eq:LSRet}$ for $\mathcal{A} = \mathcal{F}$ and $\mathcal{C} = \mathbb{C}^{\textbf{n}}$. Due to its speed and accuracy, we term our method Fast Phase Retrieval. The method achieves optimal arithmetic and sample complexity. The basis for our method is a nearly exact direct method derived from complex analysis, followed by a trust-region minimization step to remove any remaining error.  We will show that our algorithm returns a solution to \cref{PhaseRet} when $\textbf{x}$ lies in a large class of objects that we term \textit{Schwarz Objects} defined by Equation \cref{SchwarzObjects}. These vectors contain an element of dominant magnitude. However, the location of this element can be arbitrary, and its value and position are assumed unknown. The lack of position constraints or knowledge of the dominant element's value separates our method from traditional reference-based approaches \cite{barmherzig_holographic_2019}.  Fast Phase Retrieval is based on a theoretical framework that explains the numerical observations of prior work, and paves a path towards further algorithmic enhancements.

\subsection{Paper Organization}
In \cref{sec: Notation} we introduce our notation. In \cref{sec:AlgDescription} we present the mathematical framework for our approach, the resulting fast phase retrieval algorithm, and the accompanying correctness theorems. In \cref{sec: Lyapunov} we use Lyapunov methods to prove that, for a sufficiently low initial error, a local optimization will converge to a zero-cost solution, and in \cref{sec:Schwarz} we prove that the Schwarz transform achieves this error bound. 
Finally in \cref{sec:Experiments} we present numerical experiments that demonstrate the accuracy and numerical stability of our approach. 
\subsection{Mathematical Notation}\label{sec: Notation}

We follow the engineering convention by using $j$ to refer to the imaginary unit $j^2 = -1$. We use $d \in \mathbb{N}$ to denote the number of physical dimensions corresponding to a phase retrieval instance, i.e., when $\mathcal{A}$ is the d-dimensional Fourier transform. Throughout this paper, we use multi-index notation: for $\textbf{z} \in \mathbb{C}^d$ and $\textbf{k} \in \mathbb{Z}^d$ we have $\textbf{z}^\textbf{k} = \prod_{i=1}^d z_i^{k_i}$. Comparisons and arithmetic operations on multi-indices are performed element-wise. Unary operations on vectors such as modulus $|\cdot|$ or complex conjugation $\overline{\cdot}$ are also taken element-wise. 

For the vectors that arise in \cref{PhaseRet}, we refer to $\textbf{x}$ as the ``object" and $\textbf{y}$ as the ``measurement". We will treat both vectors as $d$-dimensional arrays with $\textbf{x}$ having a length along each axis given by the tuple $\textbf{n} \in \mathbb{N}^d$ and $\textbf{y}$ having lengths $\textbf{m} \in \mathbb{N}^d$. For example, a two-dimensional square object with side length 2 would have shape $\textbf{n} = (2,2)$.  
We assume complex objects and strictly positive measurements: $\textbf{x} \in \mathbb{C}^{\textbf{n}} = \mathbb{C}^{n_1 \times n_2 \times ... n_d}$ and $\textbf{y} \in \mathbb{R}^{\textbf{m}}_{>0}$. 

We denote the total object size $N = \prod_{i=1}^d n_i$ and the total measurement size $M = \prod_{i = 1}^d m_i$.

We use $\mathcal{F}$ to denote the unitary discrete Fourier transform (DFT) oversampled to $\textbf{m}$ samples with elements given by:
    \begin{equation}\label{eq:DFT}
        \mathcal{F}\{\textbf{x}\}_{\textbf{k}} =  \frac{1}{\sqrt{M}} \sum_{\textbf{i} = \textbf{0}}^{\textbf{n}-1} x_{\textbf{i}} e^{-2\pi j \textbf{k}\cdot(\textbf{i}/\textbf{m})} \quad \textbf{0} \leq \textbf{k} < \textbf{m}
    \end{equation}
We use capital bold letters to refer to the DFT of a vector $\textbf{X} = \mathcal{F}\{\textbf{x}\}$. We define the Z-Transform of $\textbf{x}$,  $X(\textbf{z}) \in \mathbb{C}[\textbf{z}]$, to be the d-variate polynomial with coefficients defined by $\textbf{x}$:
\begin{equation}
        X(\textbf{z}) = \sum_{\textbf{i}=\textbf{0}}^{\textbf{n}-\textbf{1}} x_{\textbf{i}} \textbf{z}^{\textbf{i}}
\end{equation}

We define the ``interpolation polynomial" for y to be the unique trigonometric polynomial taking on real values that interpolates the values of  $\textbf{y}$:
\begin{equation}
    y(\textbf{z}) = \frac{1}{M} |X(\textbf{z})|^2
\end{equation}
Coefficients for $y(\textbf{z})$  can be be calculated from the DFT of $\textbf{y}$.

The measurements we consider will consist of the composition of a linear operator $\mathcal{A} \in \mathcal{L}(\mathbb{C}^{\textbf{n}}, \mathbb{C}^{\textbf{m}})$ and the complex modulus $(|\cdot|)$.

We denote the poly-unit disk as $\mathbb{D}^d$  and the  poly-unit torus as $\mathbb{T}^d$, defined as:
\begin{equation}
    \mathbb{D}^d = \{\textbf{z} \in \mathbb{C}^d :|z_i| < 1 \quad \forall i \in [1,d] \}
\end{equation}
\begin{equation}
    \mathbb{T}^d = \{\textbf{z} \in \mathbb{C}^d: |z_i| = 1 \quad \forall i \in [1,d]\}
\end{equation}
Given a vector $\boldsymbol{\delta} \in \mathbb{R}_{>0}^d$, we define the corresponding annular region:
\begin{equation}
    A_{\boldsymbol{\delta}} = \{\textbf{z} \in \mathbb{C}^d : 1 - \delta_i < |z_i| < 1 + \delta_i \quad i \in [1,d] \}
\end{equation}
We denote the class of holomorphic functions over a set $\mathbb{S}$ as $\mathcal{O}(\mathbb{S})$

We will next define a multidimensional generalization of a function's winding number. In algebraic topology terms, we will be calculating the cohomology class of an algebraic curve. Our presentation will differ from standard topological texts due to the emphasis on numerical computation. 

Let $\boldsymbol{\delta} \in \mathbb{R}_{>0}^d$ be a d-tuple and $f(\textbf{z})$ be a function holomorphic on $A_{\boldsymbol{\delta}}$, non-vanishing on $\mathbb{T}^d$.  Define $f_i(z_i;\textbf{1}) $ as the one-variable function given by fixing all arguments of f to 1 except for $z_i$. We denote the derivative of $f_i$ with respect to $z_i$ as $f_i'(z_i;\textbf{1})$  We then denote its \textit{multi-winding number} around $\textbf{0}$ as $\text{Wind}(f)$. Each element of the multi-winding number $\textbf{w} = \text{Wind}(f)$ is given by

\begin{equation}
    w_i = \frac{1}{2\pi j} \oint_{\mathbb{T}} \frac{ f'_i(z_i;\textbf{1})}{f_i(z_i;\textbf{1})} d z_i
\end{equation}
For a given multi-winding number $\textbf{w}$ let $X_{-\textbf{w}}(\textbf{z})$ be the z-transform of $\textbf{x}$ with $x_{\textbf{w}}$ set to 0.  We then define the set of $\textbf{w}$-index \textit{Schwarz Objects}
\begin{equation}\label{SchwarzObjects}
    \mathbb{S}_{\textbf{w}} = \{\textbf{x} \in \mathbb{C}^\textbf{n}: \forall \textbf{z} \in \mathbb{T}^d \ \text{s.t} \ |x_\textbf{w}| \geq 2 |X_{-\textbf{w}}(\textbf{z})|\}
\end{equation}
We use $\norm{\cdot}_p$ to refer to the standard vector p-norm or the operator norm induced by the vector p-norm. We define $f_{LS}$ as the cost function for \cref{eq:LSRet} with $\mathcal{A} = \mathcal{F}$. 

\begin{equation}\label{GaussNewton}
f_{LS}(\textbf{x};\textbf{y}) =  \frac{1}{4}\norm{|\mathcal{F}\{\textbf{x}\}|^2 - \textbf{y}}_2^2
\end{equation}
To break the non-degeneracy of global phase, and to ensure the Hessian is positive definite for robustness, we define a regularized version of the sum of squares cost function.
\begin{equation}
    f_{reg}(\textbf{x};\textbf{y}, \lambda,\textbf{w}) = f_{LS}(\textbf{x};\textbf{y}) +  \frac{\lambda}{2}\Im{x_\textbf{w}}^2
\end{equation}
This regularization simply fixes the global phase such that  the $\textbf{w}$-index element of \textbf{x}, is real-valued. Finally, the \textit{gradient flow} of $f_{reg}$ is given by the following differential equation:
\begin{equation}\label{gradFlow}
    \Dot{\textbf{x}}(t) = -\nabla f_{reg}(\textbf{x}(t);\textbf{y})
\end{equation}

\section{The Fast Phase Retrieval Algorithm}\label{sec:AlgDescription}

\subsection{Mathematical Motivation}
The basis for our approach comes from a combination of traditional phase retrieval techniques, multivariate complex analysis, algebraic topology, and classical optimization.

We begin with the decades-old insight that phase retrieval is equivalent to recovering the Z-transform of an object from the magnitudes of the Z-transform on the unit circle. To see this, we observe that our problem data $\textbf{y}$ correspond to equally spaced samples of $|X(\textbf{z})|^2$ on the unit circle. By the Shannon-Nyquist  sampling theorem we can construct the unique polynomial representation of $|X(\textbf{z})|^2$ given a number of samples satisfying $\textbf{m} \geq 2 \textbf{n}$ \cite{candes_phase_2015}. Phase retrieval can then be posed as the problem of constructing a real-valued phase function $\varphi(\textbf{z})$, such that $e^{j \varphi(\textbf{z})} |X(\textbf{z})|$ is a polynomial of degree less than or equal to $\textbf{n}$. In other words, the resulting polynomial should correspond to the z-transform of an object that satisfies the support constraint. Formally, \cref{PhaseRet} can be posed as:

\begin{equation}\label{precousin}
\begin{aligned}
    \text{find} \quad &  \varphi(\textbf{z}), \textbf{x}\\
    \text{s.t.} \quad & e^{j \varphi(\textbf{z})}  |X(\textbf{z})| = \sum_{\textbf{i} = \textbf{0}}^\textbf{n} x_\textbf{i} \textbf{z}^\textbf{i}
\end{aligned}
\end{equation}
We will tackle this problem using tools from complex analysis. Concretely, we will show that \cref{precousin} can be posed as a series of two classical complex analysis problems, a continuous multiplicative Cousin problem and a Schwarz problem. Solving boundary value problems by reducing to a Cousin, Riemann-Hilbert, or Schwarz problem is a ubiquitous technique 
in numerical methods \cite{dolean_optimized_2009, its_riemann-hilbert_2003}. However, existing reductions only work for cases topologically equivalent to one complex dimension. For two or more dimensions, there are fundamental barriers that prevent the technique from being directly applied. 

However, we will show that there exists a set of \cref{PhaseRet} problems for which we can construct an exact reduction to a Cousin problem, followed by an approximate reduction to a Schwarz problem. We prove that this approximation achieves a low enough error for a local minimization to arrive at a global minimizer of \cref{PhaseRet}. Finally, we will show that there exists a simple modification that transforms any phase-retrieval problem into one that can be solved using this method.

In its simplest form, the multiplicative Cousin problem on the poly-torus is to determine zero-free smooth functions $h(\textbf{z})$ and $g(\textbf{z})$ such that for a given function $f(\textbf{z})$ smooth on $\mathbb{T}^d$ :
\begin{equation}\label{cousin}
    \frac{h(\textbf{z})}{g(\textbf{z})} =  f(\textbf{z}) \quad \forall \textbf{z} \in \mathbb{T}^d
\end{equation}
To convert $\cref{precousin}$ into this form, we set $f(\textbf{z}) = |X(\textbf{z})|$ and observe that if \cref{cousin} has a solution unique up to multiplicative constants, then we must have $h(\textbf{z}) = X(\textbf{z})$ and $g(\textbf{z}) = e^{j \varphi(\textbf{z})}$ as one solution. The traditional approach to solving \cref{cousin} is to take the logarithm of both sides:

\begin{equation}
    \log(X(\textbf{z})) = \log(|X(\textbf{z})|) + j \varphi(\textbf{z})
\end{equation}
Assuming that $X(\textbf{z})$ possesses a holomorphic logarithm, our problem is now equivalent to recovering a holomorphic function given its real part. Such a problem is known as a Schwarz problem. Formally, the Schwarz formulation of \cref{precousin} is given by:

\begin{equation}\label{Schwarz}
\begin{aligned}
\text{find} \quad & \log(X(\textbf{z})) \\
\text{s.t} \quad & \text{Re}[\log(X(\textbf{z}))] = \log(|X(\textbf{z})|) \quad \forall \textbf{z} \in \mathbb{T}^d
\end{aligned}
\end{equation}

\cref{Schwarz} can then be solved by an integral transform. In one dimension, these steps are always valid and yield the classical Hilbert transform approach to phase retrieval of stable objects. However, in two or more dimensions, this logarithmic approach is generally known to fail \cite{oka_sur_1951}. The fundamental obstruction to this approach is that if the zero set $X(\textbf{z})$ lies partially within $\mathbb{D}^d$, then $X(\textbf{z})$ will lack a holomorphic logarithm. From a topological perspective, the lack of a holomorphic logarithm is equivalent to the phase of $X(\textbf{z})$ a non-zero multiple of $2 \pi$ radians along any path on the poly-Torus. There is no way to resolve this obstruction by purely algebraic means. The classical solution from complex analysis would involve attempting to solve the highly nonlinear complex Monge-Ampere equation \cite{korevaar_several_nodate}. However, this approach is as computationally challenging as solving the original optimization problem and further will often fail as the equation will lack a solution in general. 

These obstructions make clear why this approach has not previously been pursued in multiple dimensions by the phase retrieval community. A pure mathematician would observe that there is a fundamental topological obstruction preventing the construction of an exact solution. Meanwhile, an applied mathematician attempting a numerical solution via a simple multidimensional generalization of the Hilbert transform would find results too poor to be of use. In particular, the singular nature of the relevant integral equation \cite{venugopalkrishna_fredholm_1972}.

To rescue the logarithmic approach, we will calculate and use the multi-winding index $\textbf{w}$ of $|X(\textbf{z})|$. The multi-winding is the multidimensional generalization of the classical winding number and is equivalent to the Chern class of the co-cycle represented by the function. We then apply a phase correction term $\textbf{z}^{-\textbf{w}}$ to undo the phase wrapping of $X(\textbf{z})$. Equivalently, we solve the Cousin problem for $h(\textbf{z}) = X(\textbf{z})$ and $g(\textbf{z}) = \textbf{z}^{\textbf{w}} e^{j \varphi(\textbf{z})}$. $\textbf{z}^{-\textbf{w}} X(\textbf{z})$ will have zero phase wrap and thus possess a holomorphic logarithm, allowing us to reduce the resulting Cousin problem to a Schwarz problem. However, the logarithm is holomorphic only in an annular region and not on the whole polydisc. As a consequence, the resulting Schwarz problem can no longer be exactly solved using only values on the torus. However, we will show that the values in the torus are sufficient to find an approximate solution with an error low enough that a local optimization-based search returns a true solution.

\subsection{Algorithm Description}
We denote the solution to \cref{PhaseRet} problem instance (assumed unique up to Fourier symmetries) as $\textbf{x}^{opt}$. 

Our proposed algorithm comprises 3 steps, calculating the winding number of $X(\textbf{\textbf{z}})$ from $\textbf{y}$, applying the appropriately shifted Schwarz transform, and performing a local minimization to remove the remaining error. A sufficiently accurate calculation for the winding number is given by \cref{IndexCalc}.
\begin{algorithm}[H]
    \caption{Multi-Winding Number Calculation}
    \label{IndexCalc}
\begin{algorithmic}
\STATE{Given $\textbf{y} \in \mathbb{R}_{>0}^{\textbf{m}}$} 
\STATE{Set $\textbf{c} = |\mathcal{F}^{-1}\{\textbf{y}\}|$}
\STATE{Set $\textbf{h} = \textbf{0}^{\textbf{m}}$}
\STATE{For  $\textbf{k} \leq \textbf{n}$ set $h_\textbf{k} = 1$ }
\STATE{Set $\textbf{c} = \mathcal{F}^{-1}\{\mathcal{F}\{\textbf{h}\}\odot \mathcal{F}\{\textbf{c}\}\}$ }
\STATE{Return $\textbf{w} = \arg \max \textbf{c}$}
\end{algorithmic}
\end{algorithm}
\cref{IndexCalc} convolves a simple box function with the magnitudes of the autocorrelation of $\textbf{x}^{opt}$. The $\arg \max$ returns the index where this convolution is maximum, corresponding to the element of $\textbf{x}^{opt}$ that maximizes the L-1 norm of all autocorrelation offsets including it. Since $\textbf{x}^{opt}$ is Schwarz, the largest coefficient determines $\textbf{w} = \text{Wind}(X(\textbf{z}))$ and \cref{IndexCalc} to return the true winding number.

 Given the index, we then perform the discrete index-$\textbf{w}$ Schwarz transform of $\textbf{y}$ using \cref{DST}. 
 
\begin{algorithm}[H]
    \caption{Discrete Schwarz Transform Computation}
    \label{DST}
\begin{algorithmic}
    \STATE{Given $\textbf{y} \in \mathbb{R}_{>0}^{\textbf{m}}$ and index \textbf{w}}
    \STATE{Set $\Tilde{\textbf{y}} := \mathcal{F}^{-1}\{\log(\textbf{y})\}$ }
    \STATE{Circularly shift $\Tilde{\textbf{y}}$ by \textbf{w}}
    \STATE{FOR EACH $\textbf{k}$ set $$\Tilde{\textbf{y}}_{\textbf{k}} := \begin{cases}
        \Tilde{\textbf{y}}_{\textbf{k}} & \textbf{k} = \textbf{w} \\
        2 \Tilde{\textbf{y}}_{\textbf{k}}& \textbf{k} < \textbf{m}/2, \textbf{k} \neq \textbf{w} \\
        0 & \text{otherwise}
    \end{cases}$$ }
    \STATE{Circularly shift $\Tilde{\textbf{y}}$ by $-\textbf{w}$}
    \STATE{Set $\Tilde{\textbf{y}} := \mathcal{F}\{\Tilde{\textbf{y}} \}$}
    \STATE{Return $\Tilde{\textbf{y}}$}
\end{algorithmic}
\end{algorithm}

 While the output for this algorithm is only an approximation to the discrete Schwarz transform, the error decays geometrically with the oversampling ratio. Higher accuracy can be achieved simply by resampling $\textbf{y}$ at a higher rate using its interpolation polynomial, applying \cref{DST} and downsampling back to $\textbf{m}$. However, even at the lowest sampling rate, the method is accurate to five digits, sufficient to not impact the error analysis for the rest of the algorithm. 
 
The final step of the algorithm is to run a local minimization algorithm initialized with $\Tilde{\textbf{x}}_0 = \mathcal{F}^{-1}\{\exp(\frac{1}{2} \Tilde{\textbf{y}})\}$. We will show that any reasonable local optimization method initialized with this point will converge to the unique solution to \cref{PhaseRet}. Specifically, we only require that the method be steady-state consistent with \cref{gradFlow}. We define a minimization algorithm to be steady state consistent with the gradient flow equation if its sequence of iterates $\{\textbf{s}_i\}_{i \in \mathbb{N}}$ converge to the steady state solution of \eqref{gradFlow} with initial condition $x(0) = \Tilde{x}_0$. Formally,
\begin{subequations}
\begin{equation}
    \textbf{s}_0 = \Tilde{{\textbf{x}}}_0
\end{equation}
    \begin{equation}
    \lim_{i \to \infty} \textbf{s}_i =  \lim_{t \to \infty} [\Dot{\textbf{x}}(t) = - \nabla f_{reg}(\textbf{x}(t)), \quad \textbf{x}(0) = \Tilde{\textbf{x}}_0] 
\end{equation}
\end{subequations}
To achieve superlinear convergence, we use a standard conjugate-gradient based trust region method. 

Combining all of these steps yields \textit{Fast Phase Retrieval}
\begin{algorithm}[H]
    \caption{Maskless Fast Phase Retrieval}
    \label{MasklessFastPhase}
\begin{algorithmic}
    \STATE{Input $\textbf{y} \in \mathbb{R}_{>0}^{\textbf{m}}$, support $\textbf{n}$,and tolerance $\epsilon$}
    \STATE{Set $\textbf{w} = \text{Ind}_X$ \quad (\cref{IndexCalc})}
    \STATE{Set $\textbf{x}_0 :=  \mathcal{F}\{ \exp(\frac{1}{2} \mathcal{S}_{\textbf{w}}\{\textbf{y}\})\}$ \quad (\cref{DST})}
    \STATE{Set $\textbf{x}_1 := \textbf{TrustRegion}(f_{reg}, \textbf{x}_0, \epsilon) $}
    \RETURN{$\textbf{x}_1$}
\end{algorithmic}
    
\end{algorithm}

\subsection{Proof of correctness}
We next provide the proof of correctness. We begin by defining the desired basin of attraction for the gradient flow equation. 

\begin{theorem}[Exponential Stability Region of the Gradient Flow Equation]
    \label{gradStability}
    Given vectors $\Tilde{\textbf{x}}_0 \in \mathbb{C}^{\textbf{n}}$,$\textbf{y} \in \mathbb{C}^\textbf{m}$, and $\textbf{w} \in \mathbb{N}_0^d$ define
    \begin{equation}
        \textbf{x}^{opt} \equiv \arg \min_{\textbf{x}} f_{reg}(\textbf{x};\textbf{y},1,\textbf{w})
    \end{equation}
    \begin{equation}
        \boldsymbol{\epsilon} \equiv \textbf{x}^{opt} - \Tilde{\textbf{x}}_0
    \end{equation}
    If $\norm{\mathcal{F}\{ \boldsymbol{\epsilon}\}}_4^4 < |\mathcal{F}\{ \boldsymbol{\epsilon}\}|^{2T} |\Re{\mathcal{F}\{ \boldsymbol{\epsilon}\} \mathcal{F}\{\textbf{x}^{opt}\}}|$ then the solution $\textbf{x}(t)$ to \eqref{gradFlow} initialized with $\textbf{x}(0) = \Tilde{\textbf{x}}_0$ converges exponentially to $\textbf{x}^{opt}$:
\begin{subequations}
    \begin{equation}
        \lim_{t \to \infty} \textbf{x}(t) = \textbf{x}^{opt}
    \end{equation}
    \begin{equation}
        \exists \alpha,\beta \in \mathbb{R}_{<0} \ :\ \norm{\textbf{x} - \textbf{x}^{opt}} \leq \alpha \norm{\textbf{x}(0)\ - \textbf{x}^{opt}} e^{-\beta t}\quad \forall t>0
    \end{equation}
\end{subequations}
\end{theorem}
\begin{proof}
    See \cref{sec: Lyapunov}
\end{proof}
\begin{theorem}[Approximate Discrete Schwarz Transform Error Bound]
    If $\textbf{x}^{opt}$ is a Schwarz object then the output of \cref{DST} satisfies the requirements of \cref{gradStability}.
\end{theorem}
\begin{proof}
    See \cref{sec:Schwarz}
\end{proof}

\begin{theorem}[Fast Phase Retrieval Correctness]
    Assume that there exists a $\textbf{w}$-index Schwarz Object that solves \eqref{PhaseRet}. Then the minimization step of \cref{MasklessFastPhase} will terminate after finitely many steps bounded by $\Tilde{O}( \log(\frac{1}{\epsilon}))$, the required number of arithmetic operations will be bounded by $\Tilde{O}(N \log(N))$ and the algorithm output $\Tilde{x}$ will satisfy 
    \begin{equation}
        f_{reg}(\Tilde{\textbf{x}};\textbf{y},1, \textbf{w}) \leq \epsilon 
    \end{equation}
\end{theorem}
\begin{proof}
\cref{gradStability} guarantees that a gradient-based method initialized with $\Tilde{\textbf{x}}_0$ is guaranteed to achieve at least linear convergence \cite{moucer_systematic_2023}, giving us our logarithmic error bound. Since gradient and conjugate gradient calculations can be computed in $\Tilde{O}(N \log(N))$ time, and the required number of steps is independent of problem size, we also achieve the desired arithmetic complexity of $\Tilde{O}(N \log(N))$. 

\end{proof}

\subsection{Masked Fast Phase Retrieval}
While the previous algorithm is limited in its application to $\mathcal{A} = \mathcal{F}$. A sequence of two measurements can be used with the algorithm to recover any object.

\begin{equation}
\begin{aligned}
    \mathcal{A}_1 = & \mathcal{I}\\\
    \mathcal{A}_2 = & \mathcal{F}\mathcal{D}_{\textbf{w}}[|\textbf{x}|]
\end{aligned}
\end{equation}

Where $\mathcal{I}$  is the identity operator and $\mathcal{D}_{\textbf{k}} = r^{-|\textbf{k} - \textbf{n}|_1}$  is a diagonal operator with exponentially decaying coefficients.  With a sufficiently small base $r$, the object becomes a Schwarz object. Applying \cref{MasklessFastPhase}, and then applying the inverse $\mathcal{D}$ successfully the complex vector. Although this measurement is far from optimal, it does demonstrate such a measurement exists for any complex vector.

\section{Asymptotic Stability of Gradient Flow Equation by Lyapunov \hfill \break Analysis} \label{sec: Lyapunov}
We will use the tools of Lyapunov analysis \cite{moucer_systematic_2023} to prove \cref{gradStability}. We prove convergence via Lyapunov analysis since the cost function is non-convex, and its Hessian will generally have negative eigenvalues at the initial point.
Before we begin, we require the following lemma, similar to the positivestellensatz \cite{parrilo_chapter_2012}. :
\begin{lemma}\label{postives}
    Let $g(\textbf{x})$ be a continuous function and W be the set defined by $W = \{\textbf{x} \in \mathbb{R}^N: \textbf{g}(x) < \textbf{0} \}$. Let $\Sigma^2[\textbf{X}]$ be the set of functions expressible as the sum of squares of algebraic functions of indeterminate $\boldsymbol{X}$. Let $p(\textbf{x})$ be another continuous function. If the following holds:
    \begin{equation}
        p(\textbf{x}) +  \textbf{g}(\textbf{x}) \in \Sigma^2[\textbf{X}]
    \end{equation}
    Then $p(\textbf{x}) > 0$ for all $ \textbf{x} \in W$
\end{lemma}
\begin{proof}
    A sum of squares is trivially non-negative. We thus have in the set W:
    \begin{equation}
        p(\textbf{x}) +  g(\textbf{x}) \geq 0 \Rightarrow p(\textbf{x}) \geq -  g(\textbf{x}) > 0 \Rightarrow p(\textbf{x}) > 0
    \end{equation}
\end{proof}
With this lemma in hand, we prove the main theorem:
\begin{theorem}[Exponential Stability Region of Gradient Flow Equation]
    Let $\textbf{x}_*$ be a point satisfying $f_{LS}(\textbf{x}_*) = 0$, $\boldsymbol{\epsilon}$ be a point in error coordinates defined by  $\boldsymbol{\epsilon}(\textbf{x}) = \textbf{x} - \textbf{x}_*$, and define a candidate Lyapunov function $V(\boldsymbol{\epsilon}) = \frac{1}{2} \norm{\boldsymbol{\epsilon}}$. Then for all points satisfying $g(\boldsymbol{\epsilon}) = \norm{\mathcal{F}\{ \boldsymbol{\epsilon}\}}_4^4 - \langle |\mathcal{F}\{ \boldsymbol{\epsilon}\}|^2,  |\Re{\mathcal{F}\{ \boldsymbol{\epsilon}\} \overline{\mathcal{F}\{\textbf{x}_*\}}}| \rangle < 0$, the following holds:
    \begin{enumerate}
        \item  $V(\boldsymbol{\epsilon}) \geq 0$
        \item $V(0) \Leftrightarrow \boldsymbol{\epsilon} = 0$
        \item $\Dot{V}(\boldsymbol{\epsilon}) \equiv  \langle -\nabla f_{reg}, \nabla V(\boldsymbol{\epsilon}) \rangle  < 0$
    \end{enumerate}
    Additionally, define arbitrary constants $0 < \delta < 1$ and $\alpha > 0$. Then for all $\boldsymbol{\epsilon}$ that satisfy $\norm{\mathcal{F}\{ \boldsymbol{\epsilon}\}}_4^4 < \delta \langle |\mathcal{F}\{ \boldsymbol{\epsilon}\}|^2,  |\Re{\mathcal{F}\{ \boldsymbol{\epsilon}\} \mathcal{F}\{\textbf{x}_*\}}| \rangle$ and $\norm{\text{Re}\left[ \mathcal{F}\{\boldsymbol{\epsilon}\} \odot \overline{\mathcal{F}\{\textbf{x}_*\}} \right]}^2 > \alpha \norm{\boldsymbol{\epsilon}}^2$ the following holds:
    
    \begin{enumerate}[resume*]
        \item $\Dot{V}(\boldsymbol{\epsilon}) \leq -\frac{(1-\delta)(7+\delta)}{4} \alpha V(\boldsymbol{\epsilon})$
    \end{enumerate}
\end{theorem}
\begin{proof}
We work with Wirtinger gradients for ease of notation. Points 1 and 2 are trivially true. To prove point 3 we begin with gradient calculations.  For   $V(\boldsymbol{\epsilon})$ and $f_{reg}(\boldsymbol{\epsilon})$ the gradients are as follows:
\begin{equation}
    \nabla V(\boldsymbol{\epsilon}) = \boldsymbol{\epsilon}
\end{equation}
\begin{equation}
    \nabla f(\boldsymbol{\epsilon}) = \mathcal{F}^{\dagger}\{\left[|\mathcal{F}\{ \boldsymbol{\epsilon}\}|^2 + 2 \text{Re}[\mathcal{F}\{ \boldsymbol{\epsilon}\} \odot \overline{\mathcal{F}\{ \textbf{x}_*\}}] \right] \odot \mathcal{F}\{\boldsymbol{\epsilon} + \textbf{x}_*\} \} + 1j*\Im[\boldsymbol{\epsilon}_\textbf{w}] \
\end{equation}
For ease of notation let $\boldsymbol{\mathcal{E}} = \mathcal{F}\{\boldsymbol{\epsilon}\}$ and $\textbf{X}_* = \mathcal{F}\{\textbf{x}_*\}$. 
We then have:
\begin{equation}
\begin{aligned}
    -\Dot{V}(\textbf{x}) &= \Re{(\nabla V(\textbf{x}))^\dagger \nabla f(\textbf{x})}  \\
    & =  \norm{\boldsymbol{\mathcal{E}}}_4^4  + 3 |\boldsymbol{\mathcal{E}}|^{2T} \Re[\boldsymbol{\mathcal{E}} \odot  \overline{\textbf{X}_*}] + 2 \norm{\Re[\boldsymbol{\mathcal{E}} \odot  \overline{\textbf{X}_*}]}_2^2 + \Im[\boldsymbol{\epsilon}_\textbf{w}]^2 \\
    & \geq  \norm{\boldsymbol{\mathcal{E}}}_4^4  - 3 |\boldsymbol{\mathcal{E}}|^{2T} |\Re[\boldsymbol{\mathcal{E}} \odot  \overline{\textbf{X}_*}]| + 2 \norm{\Re[\boldsymbol{\mathcal{E}} \odot  \overline{\textbf{X}_*}]}_2^2
\end{aligned}
\end{equation}
We prove this expression is positive (and thus $\Dot{V}$ is negative) via \cref{postives}. 

In our case, the indeterminates are the real and imaginary parts of $\boldsymbol{\epsilon}$, and $p$ corresponds to the expression above viewed as a function of real variables. We then let $g$ be $\norm{\boldsymbol{\mathcal{E}}}_4^4 - \langle |\boldsymbol{\mathcal{E}}|^2,  |\Re[\boldsymbol{\mathcal{E}} \odot \textbf{X}_*]| \rangle$ viewed as a real function. We then have:
\begin{equation}
    p + \norm{\boldsymbol{\mathcal{E}}}_4^4 - \langle |\boldsymbol{\mathcal{E}}|^2,  |\Re[\boldsymbol{\mathcal{E}} \odot \textbf{X}_*]| \rangle = 2 \norm{ |\boldsymbol{\mathcal{E}}|^2 - | |\Re[\boldsymbol{\mathcal{E}} \odot \textbf{X}_*]|}_2^2 \in \Sigma^2
\end{equation}

Invoking \cref{postives} then yields claim 3. 
To prove claim 4 we modify the above to $p + \norm{\boldsymbol{\mathcal{E}}}_4^4 - \delta \langle |\boldsymbol{\mathcal{E}}|^2,  |\Re[\boldsymbol{\mathcal{E}} \odot \textbf{X}_*]| \rangle $:
\begin{equation}
\begin{aligned}
        \norm{ \sqrt{2}|\boldsymbol{\mathcal{E}}|^2 - (\sqrt{2} - \frac{1 - \delta}{2 \sqrt{2}})| |\Re[\boldsymbol{\mathcal{E}} \odot \textbf{X}_*]|}_2^2 
        +\frac{(1-\delta)(7 + \delta)}{8}  \norm{\Re[\boldsymbol{\mathcal{E}} \odot  \overline{\textbf{X}_*}]}_2^2 \\
        \geq \norm{ \sqrt{2}|\boldsymbol{\mathcal{E}}|^2 - (\sqrt{2} - \frac{1 - \delta}{2 \sqrt{2}})| |\Re[\boldsymbol{\mathcal{E}} \odot \textbf{X}_*]|}_2^2 + \frac{(1-\delta)(7+\delta)}{4} \alpha V(\epsilon)
\end{aligned}
\end{equation}
Invoking the non-negativity norms proves claim 4 directly. 
\end{proof}
From these claims, exponential convergence of the continuous time model follows from standard Lyapunov theory.

\section{Schwarz Transform Error Bound}
\label{sec:Schwarz}
\subsection{Generalizing the Schwarz Integral}
The traditional solution to the Schwarz problem is given by the following integral transform
\begin{equation}
\label{eq: Schwarz Transform}
    y(\textbf{z}) = \frac{1}{(2\pi j)^d}\int_{\mathbb{T}^d} \log(X(\textbf{z})) \left[2 \prod_{i =1}^d \frac{1}{1 - z_i/\zeta_i} - 1\right] \frac{d \boldsymbol{\zeta}}{\boldsymbol{\zeta}} + j \varphi \quad \forall \textbf{z} \in \Delta_d,\  \varphi \in \mathbb{R}
\end{equation}

With $\varphi$ representing a free real constant. This integral transform fully captures the set of symmetries for \cref{PhaseRet}. The free constant corresponds to a change of global phase, changing the orientation of the integration contour corresponds to performing a conjugate flip, and the choice in the logarithm branch corresponds to circular shifts. In 1-dimension \cref{eq: Schwarz Transform} is simply the Hilbert transform method, while in multiple dimensions the Schwarz integral does not correspond to any of the common multidimensional generalizations of the Hilbert transform. The existence of a direct method for solving \cref{PhaseRet} in multiple dimensions appears to be unknown in the phase-retrieval literature, making its discovery (even for the limited cases in which it applies) an important milestone. 

In order to make use of the transform, we need to tackle two obstacles. First, we need to define a logarithm holomorphic on an annular region; a global logarithm cannot be used since f vanishes in general. We will refer to the defined logarithm as the \textit{Annular Logarithm}. Second, we have to show that applying the Schwarz transform to the Annular Logarithm results in an error that satisfies the requirements \cref{gradStability}.

\begin{theorem}[Annular Logarithm Existence]
For all f non-zero and holomorphic on Annular region $A_\epsilon^d$, there exists $\textbf{w} \in \mathbb{Z}^d$ and $g(\textbf{z})$ holomorphic on this region such that for all $\textbf{z}, \textbf{z}_0 \in A_{\epsilon}^d$:
\begin{equation}
\begin{aligned}
    f(\textbf{z}) &= \textbf{z}^{\textbf{-w}}e^{g(\textbf{z})} \\
    \textbf{w} &= \text{Wind}(f)\\
    g(\textbf{z}) &= \log(\textbf{z}_0^{-\textbf{w}} f(\textbf{z}_0)) + \int_{\textbf{z}_0}^\textbf{z} \frac{\partial\left[\boldsymbol{\zeta}^{- \textbf{w}} f(\boldsymbol{\zeta}) \right]}{\boldsymbol{\zeta}^{-\textbf{w}} f(\boldsymbol{\zeta})} d \boldsymbol{\zeta}\\
\end{aligned}
\end{equation}
\end{theorem}
\begin{proof}
The proof in one complex dimension is a common exercise. However, extension to multiple dimensions requires a modification. We first recall the definition of a winding number, treating each element as a function of $\textbf{z}_i'$, the variable $\textbf{z}$ with $z_i$ removed:
\begin{equation}
    w_i(\textbf{z}_i') = \frac{1}{2\pi j} \oint_{\mathbb{T}} \frac{ f'_i(z_i;\textbf{z}_i')}{f_i(z_i;\textbf{z}_i')} d z_i
\end{equation}
We observe that $\frac{f_i'(z_i;\textbf{z}_i')}{f_i(z_i;\textbf{z}_i')}$ is holomorphic with respect to $\textbf{z}_i'$ since it is the quotient of two holomorphic functions with non-vanishing denominator. Thus, $w_i(\textbf{z}_i')$ is holomorphic. Next, we observe that the winding number takes only integer values. $w_i(\textbf{z}_i')$ is thus both continuous and integer-valued and therefore must be constant. By the identity theorem, $w_i(\textbf{z}_i')$ will then be constant for all $\textbf{z}_i'$ on the Torus. Finally, we observe that since $f(\textbf{z})$ is non-vanishing on the Torus, it will have a constant winding number of 0 for any path contractible to a point on the Torus. 

Finally, combining all previous observations, we conclude that $\textbf{z}^{-\textbf{w}} f(\textbf{z})$ will have zero winding number for all loops on the Torus; the loop will be contractible on the Torus and vanish or enclose an equal number of poles and zeros and vanish by the argument principle. This allows the local definition of a logarithm.

\end{proof}
\subsection{Proving Error Bound of Schwarz Transform}
The logarithm for a $\textbf{w}$-index Schwarz object can be expressed as:
\begin{equation}
    \log(\textbf{z}^{-\textbf{w}} X(\textbf{z})) = \log(x_{\textbf{w}} + X_{- \textbf{w}}(\textbf{z})) = \log(x_{\textbf{w}}) + \log(1 + \frac{X_{- \textbf{w}}(\textbf{z})}{x_{\textbf{w}}}) 
\end{equation}
Examining the action of the Schwarz transform on the real part of this expression allows us to prove the error bound. The Schwarz transform functionally truncates the Laurent series for the logarithm at $\textbf{k} = - \textbf{w}$ and then scales all entries not at the origin by 2. To calculate the error, we begin by expressing the Laurent series of the logarithm (this series is guaranteed to be absolutely convergent by the definition of Schwarz objects). 

\begin{equation}
    \log(1 + \frac{X_{- \textbf{w}}(\textbf{z})}{x_{\textbf{w}}})  = \sum_{k=1}^\infty \frac{(-1)^{k+1}}{k} \left(\frac{X_{- \textbf{w}}(\textbf{z})}{x_{\textbf{w}}}\right)^k
\end{equation}
Taking the real part performs a conjugate flip about the origin. Equivalently, we add the conjugate expression. We then multiply by 2 to account for the Schwarz transform:
\begin{equation}
    2\Re{\log(1 + \frac{X_{- \textbf{w}}(\textbf{z})}{x_{\textbf{w}}})} = \sum_{k=1}^\infty \frac{(-1)^{k+1}}{k} \left(\frac{X_{- \textbf{w}}(\textbf{z})}{x_{\textbf{w}}}\right)^k + \frac{(-1)^{k+1}}{k} \left(\frac{X^\dagger_{- \textbf{w}}(\textbf{z})}{x_{\textbf{w}}}\right)^k
\end{equation}
\begin{equation}
    = \log(1 + \frac{X_{- \textbf{w}}(\textbf{z})}{x_{\textbf{w}}} + \frac{X^\dagger_{- \textbf{w}}(\textbf{z})}{x_{\textbf{w}}})
\end{equation}

We can then find the output of the exponential of the Schwarz transform:

\begin{equation}
    \exp{\mathcal{S}_\textbf{w}\{|\log(\textbf{z}^{-\textbf{w}} X(\textbf{z}))|\}} = X(\textbf{z})+ X^\dagger_{- \textbf{w}}(\textbf{z})
\end{equation}
Where $X^\dagger$ is the polynomial with coefficients reversed and conjugated. 
The Fourier transform of the error is then:
\begin{equation}
    \mathcal{F}\{\boldsymbol{\epsilon}\} = \mathcal{F}\{\overline{\textbf{x}^{opt}}\} - x_\textbf{w}
\end{equation}
From the definition of Schwarz objects, we always have $\Re{(\overline{\mathcal{F}\{\textbf{x}^{opt}\}})^2} > 0$, and $\Re{\overline{\mathcal{F}\{\textbf{x}^{opt}}\}} > 0$. Therefore, we have the inequality:

\begin{equation}
    | \overline{\mathcal{F}\{\textbf{x}^{opt}\}} - x_\textbf{w}|^2 \leq \Re{\overline{\textbf{y}} (\textbf{x}_{\textbf{w}} - \mathcal{F}\{\textbf{x}^{opt}\})} = |\Re[\mathcal{F}\{\boldsymbol{\epsilon}\} \odot \overline{\mathcal{F}\{\textbf{x}^{opt}\}}]|
\end{equation}

Combining this with Cauchy-Schwarz implies that the output of the Schwarz transform lies in the basin of attraction of the gradient flow equation.

\section{Numerical Experiments} \label{sec:Experiments}
We perform several numerical experiments to verify the efficiency and noise resilience of Fast Phase Retrieval. We implemented the algorithm using Pytorch, making use of GPU acceleration. All experiments were run on a commercially available laptop with an NVIDIA GTX 3070.

\subsection{Numerical Evaluation of the Schwarz Transform}
\begin{figure}[H]
    \centering
    \includegraphics[scale = .5]{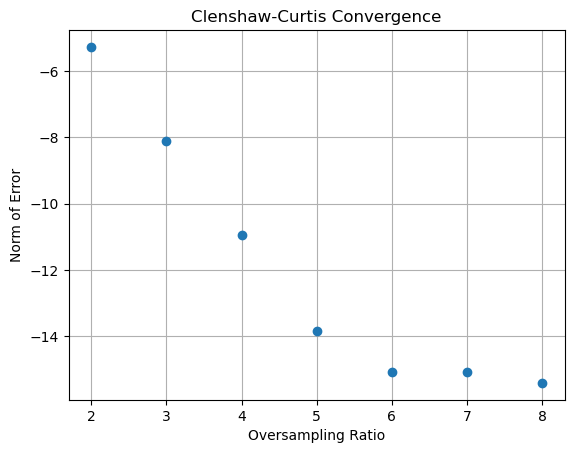}
    \caption{Relative Error of Clenshaw-Curtis Quadrature for the Evaluation of the Schwarz Transform} \label{fig:clenshawCurtis}
\end{figure}

The continuous integral definition of the Schwarz transform is not tractable for numerical evaluation. To numerically approximate the transform, we observe that the integral is equivalent to applying a mask to the Fourier coefficients of $\log(\textbf{z}^{-\textbf{w}} y(\textbf{z}))$. Since this function is analytic in a region containing the poly-Torus, its Fourier coefficients will decay geometrically. Combining this observation with the fact that we already have sampled the function at roots of unity and require the integral values at roots of unity makes the Clenshaw-Curtis quadrature \cite{trefethen_is_2008} the obvious approach. \cref{fig:clenshawCurtis} shows the error vs. oversampling factor, and geometric convergence is clearly demonstrated. Even at a critical oversampling of $\textbf{m} = 2 \textbf{n}$ a relative error below $10^{-5}$, below any level that would impact convergence of the rest of the algorithm.

\subsection{Gaussian Noise Experiments}
To test Fast Phase Retrieval's robustness against noise, we generated problem instances simulating Gaussian noise. For each test, we generate a 128x128 complex vector consisting of i.i.d. standard complex normal variables. We then oversample the Fourier magnitudes of this object by a factor and add Gaussian noise corresponding to different SNR levels. In decibels, the signal-to-noise ratio is given by SNR $ = 10 \log_{10} \frac{\norm{x}^2}{\sigma^2} $ where $\sigma^2$ is the corresponding Gaussian variance,  We report RMSE as $10 \log_{10} \frac{\norm{\hat{x} - \textbf{x}_{true}}}{\norm{\textbf{x}_{true}}}$. To improve the numerical convergence of local minimization methods, we replace \cref{eq:LSRet} with \cref{eq:MaxLikRet}. The substitution corresponds to normalizing with the maximum likelihood cost function for the Gaussian approximation to Poissonian noise, which improves the numerical convergence. 
\begin{equation}\label{eq:MaxLikRet}
    \min_{\textbf{x} \in \mathcal{C}} \quad \frac{1}{8}\norm{\frac{1}{\sqrt{\textbf{y}}} \odot (|\mathcal{A}\{\textbf{x}\}|^2 - \textbf{y})}_2^2 
\end{equation}
We tested two different classes of object, one with a zero winding number, and one without a guaranteed zero-winding number. To generate a zero winding number, we place an impulse of brightness $128^2$ at the origin of the object. For the curve labelled non-zero winding number, this impulse is of brightness $128^2$ located at position (1,1). Importantly, this reference location violates the holographic separation principle, and the location of the reference is not given as input to the algorithm.  For each SNR level, we ran 100 experiments for each object. It took 60.2 seconds in total to run all 1200 recoveries.

\begin{figure}[H]
    \centering
    \includegraphics[scale = .75]{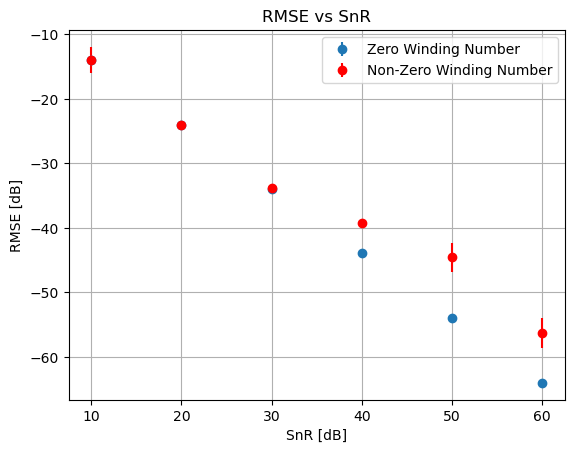}
    \caption{Comparison of RMSE vs SnR for Different Reference Spot Location. Reference bars are minimum vs maximum observed RMSE.}
    \label{fig:SNR}
\end{figure}
 
 Figure \ref{fig:SNR} shows a comparison of the RMSE of these two categories as the SNR changes. We see that for relatively large noise values, the RMSE scaling is identical for the two reference point positions. However, as the point moves inside the object, the scaling differs based on the reference point location. This difference in scaling likely arises from the exactness of the Schwarz transform when the impulse lies on the boundary of the object. 

\begin{figure}[H]
    \centering
    \includegraphics[scale = .75]{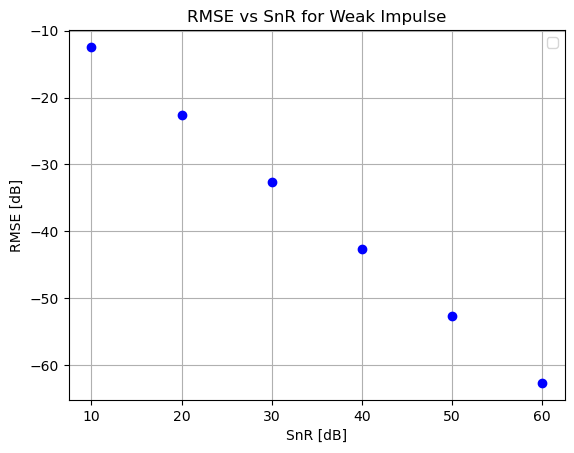}
    \caption{RMSE vs SNR for Weak Impulse. }
    \label{fig:weakimpulse}
\end{figure}

 To demonstrate the performance of our algorithm on more reasonable objects, we repeated the experiment on another object with a bright spot of magnitude $925 = 128*5$ (corresponding to $4.2\%$ of the 1-Norm of the object) located at position (0,0). Plotted in \cref{fig:weakimpulse} we observe identical RMSE scaling, demonstrating that fast phase retrieval succeeds even for a much weaker bright point.
\subsection{Numerical comparison to state-of-the-art}
\begin{figure}[H]
     \centering
    \includegraphics[scale = .5]{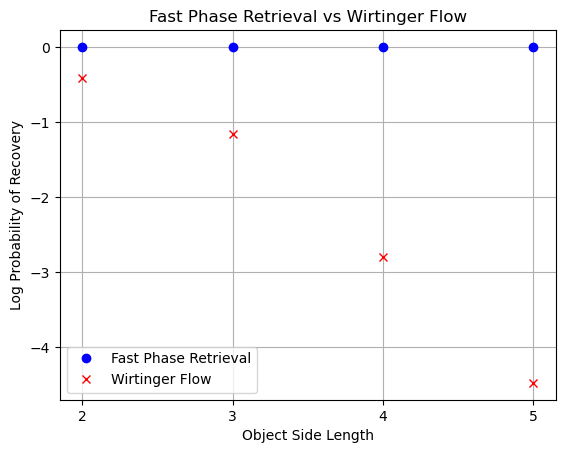}
     \caption{Wirtinger Flow vs Fast Phase Retrieval}
     \label{fig:WirtingervsFast}
 \end{figure}

 In \cref{fig:WirtingervsFast} we present numerical experiments that demonstrate the superior convergence properties of Fast Phase Retrieval compared to Wirtinger Flow, the standard gradient-based phase retrieval algorithm. Since the most popular \cite{candes_phase_2015} initializations of Wirtinger Flow failed to converge, we instead chose random initial guess taken from a complex Gaussian distribution. Even for square objects with a side length of 5, we see that the probability of a random initialization succeeding is vanishingly small. $10^5$ initializations were used for each side length.

\section{Conclusions}
\label{sec:conclusions}

Fast Phase Retrieval serves as the first and only algorithm capable of achieving deterministic recovery of complex objects from their oversampled Fourier transform in a polynomial number of arithmetic operations. Our algorithm achieves this recovery without requiring any manipulation of the object (such as adding a reference beam) nor requiring additional non-Fourier measurements, overcoming the strict limitations of existing algorithms. This is crucial for experimental settings such as noncrystallographic x-ray imaging or quantum state tomography, where modification of either the object or the measurement vector is challenging. Furthermore, the stability to noise, optimal arithmetic complexity - $\Tilde{O}(N \log(N))$ - and optimal sample complexity - $\Tilde{O}(N)$ -  make it a practical tool for real-life settings, time-critical applications such as in a feedback loop, and data heavy processing such as with mega-pixel camera apertures.

We hope that the success of our mathematical method motivates further investigation into the practical application of modern algebraic and topological results in inverse problems. It will also be exciting to see if the numerical results presented could aid algebraic topologists in practical problem solving. Furthermore, this approach can be used for tackling previously unsolved instances of non-convex quadratically constrained quadratic programs. 

Finally, we emphasize that our results finally prove the long-standing conjecture that oversampled Fourier measurements are sufficient to solve the phase-retrieval problem stably and efficiently, a conjecture which had been cast into doubt due to the shortcomings of previous approaches \cite{candes_phase_2015}. 

\section*{Acknowledgments}
We thank Prof. Yonina Eldar, Prof. Laurent Demanet, Prof. Jianwei Miao and Prof. Alexandre Megretski for useful discussions.
The authors also thank Prof. Mohamed ElKabbash, Thomas Propson, and Ian Christen.

STM acknowledges the support of the MITRE Quantum Moonshot Program.
An early version of this work appeared as part of CB's Master Thesis "Fast Phase Retrieval: A Robust and Efficient Multidimensional Phase Retrieval Algorithm".

\bibliographystyle{siamplain}
\bibliography{fastPhase_new}   

\begin{thebibliography}{10}

\bibitem{arnon_cylindrical_1984}
{\sc D.~S. Arnon, G.~E. Collins, and S.~McCallum}, {\em Cylindrical {Algebraic} {Decomposition} {I}: {The} {Basic} {Algorithm}}, SIAM Journal on Computing, 13 (1984), pp.~865--877, \url{https://doi.org/10.1137/0213054}, \url{https://epubs.siam.org/doi/abs/10.1137/0213054} (accessed 2024-01-26).

\bibitem{barmherzig_holographic_2019}
{\sc D.~A. Barmherzig, J.~Sun, P.-N. Li, T.~J. Lane, and E.~J. Candès}, {\em Holographic phase retrieval and reference design}, Inverse Problems, 35 (2019), p.~094001, \url{https://doi.org/10.1088/1361-6420/ab23d1}, \url{https://dx.doi.org/10.1088/1361-6420/ab23d1} (accessed 2024-06-08).
\newblock Publisher: IOP Publishing.

\bibitem{bendory_toward_2020-1}
{\sc T.~Bendory and D.~Edidin}, {\em Toward a {Mathematical} {Theory} of the {Crystallographic} {Phase} {Retrieval} {Problem}}, SIAM Journal on Mathematics of Data Science, 2 (2020), pp.~809--839, \url{https://doi.org/10.1137/20M132136X}, \url{https://epubs.siam.org/doi/abs/10.1137/20M132136X} (accessed 2024-07-01).
\newblock Publisher: Society for Industrial and Applied Mathematics.

\bibitem{cai_optimal_2016}
{\sc T.~T. Cai, X.~Li, and Z.~Ma}, {\em Optimal rates of convergence for noisy sparse phase retrieval via thresholded {Wirtinger} flow}, The Annals of Statistics, 44 (2016), pp.~2221--2251, \url{https://doi.org/10.1214/16-AOS1443}, \url{https://projecteuclid.org/journals/annals-of-statistics/volume-44/issue-5/Optimal-rates-of-convergence-for-noisy-sparse-phase-retrieval-via/10.1214/16-AOS1443.full} (accessed 2024-04-16).
\newblock Publisher: Institute of Mathematical Statistics.

\bibitem{candes_phase_2015}
{\sc E.~J. Candès, Y.~C. Eldar, T.~Strohmer, and V.~Voroninski}, {\em Phase {Retrieval} via {Matrix} {Completion}}, SIAM Review, 57 (2015), pp.~225--251, \url{https://doi.org/10.1137/151005099}, \url{http://epubs.siam.org/doi/10.1137/151005099} (accessed 2023-09-13).

\bibitem{candes_solving_2014}
{\sc E.~J. Candès and X.~Li}, {\em Solving {Quadratic} {Equations} via {PhaseLift} {When} {There} {Are} {About} as {Many} {Equations} as {Unknowns}}, Foundations of Computational Mathematics, 14 (2014), pp.~1017--1026, \url{https://doi.org/10.1007/s10208-013-9162-z}, \url{https://doi.org/10.1007/s10208-013-9162-z} (accessed 2024-04-16).

\bibitem{candes_phaselift_2013}
{\sc E.~J. Candès, T.~Strohmer, and V.~Voroninski}, {\em {PhaseLift}: {Exact} and {Stable} {Signal} {Recovery} from {Magnitude} {Measurements} via {Convex} {Programming}}, Communications on Pure and Applied Mathematics, 66 (2013), pp.~1241--1274, \url{https://doi.org/10.1002/cpa.21432}, \url{https://onlinelibrary.wiley.com/doi/abs/10.1002/cpa.21432} (accessed 2024-04-16).
\newblock \_eprint: https://onlinelibrary.wiley.com/doi/pdf/10.1002/cpa.21432.

\bibitem{dolean_optimized_2009}
{\sc V.~Dolean, M.~Gander, and L.~Gerardo-Giorda}, {\em Optimized {Schwarz} {Methods} for {Maxwell}'s {Equations}}, SIAM Journal on Scientific Computing, 31 (2009), pp.~2193--2213, \url{https://doi.org/10.1137/080728536}, \url{https://epubs.siam.org/doi/abs/10.1137/080728536} (accessed 2024-04-16).
\newblock Publisher: Society for Industrial and Applied Mathematics.

\bibitem{fienup_phase_1982}
{\sc J.~R. Fienup}, {\em Phase retrieval algorithms: a comparison}, Applied Optics, 21 (1982), pp.~2758--2769, \url{https://doi.org/10.1364/AO.21.002758}, \url{https://opg.optica.org/ao/abstract.cfm?uri=ao-21-15-2758} (accessed 2024-03-19).
\newblock Publisher: Optica Publishing Group.

\bibitem{goldstein_phasemax_2018}
{\sc T.~Goldstein and C.~Studer}, {\em {PhaseMax}: {Convex} {Phase} {Retrieval} via {Basis} {Pursuit}}, IEEE Transactions on Information Theory, 64 (2018), pp.~2675--2689, \url{https://doi.org/10.1109/TIT.2018.2800768}, \url{https://ieeexplore.ieee.org/abstract/document/8278279} (accessed 2024-04-16).
\newblock Conference Name: IEEE Transactions on Information Theory.

\bibitem{its_riemann-hilbert_2003}
{\sc A.~Its}, {\em The {Riemann}-{Hilbert} {Problem} and {Integrable} {Systems}}, 2003, \url{https://www.semanticscholar.org/paper/The-Riemann-Hilbert-Problem-and-Integrable-Systems-Its/e714066bc17b5d84ff0cff0207a6e7bdbc1099ef} (accessed 2024-04-16).

\bibitem{jaganathan_phase_2016}
{\sc K.~Jaganathan, Y.~C. Eldar, and B.~Hassibi}, {\em Phase {Retrieval}: {An} {Overview} of {Recent} {Developments}}, in Optical {Compressive} {Imaging}, CRC Press, 2016.
\newblock Num Pages: 34.

\bibitem{josz_lasserre_2018}
{\sc C.~Josz and D.~K. Molzahn}, {\em Lasserre {Hierarchy} for {Large} {Scale} {Polynomial} {Optimization} in {Real} and {Complex} {Variables}}, SIAM Journal on Optimization, 28 (2018), pp.~1017--1048, \url{https://doi.org/10.1137/15M1034386}, \url{https://epubs.siam.org/doi/abs/10.1137/15M1034386} (accessed 2024-01-26).

\bibitem{korevaar_several_nodate}
{\sc J.~Korevaar and J.~Wiegerinck}, {\em Several {Complex} {Variables}}.

\bibitem{moucer_systematic_2023}
{\sc C.~Moucer, A.~Taylor, and F.~Bach}, {\em A {Systematic} {Approach} to {Lyapunov} {Analyses} of {Continuous}-{Time} {Models} in {Convex} {Optimization}}, SIAM Journal on Optimization, 33 (2023), pp.~1558--1586, \url{https://doi.org/10.1137/22M1498486}, \url{https://epubs.siam.org/doi/full/10.1137/22M1498486} (accessed 2024-01-26).

\bibitem{oka_sur_1951}
{\sc K.~Oka}, {\em Sur les {Fonctions} {Analytiques} de {Plusieurs} {Variables}, {VIII}--{Lemme} {Fondamental} ({Suite})}, Journal of the Mathematical Society of Japan, 3 (1951), pp.~259--278, \url{https://doi.org/10.2969/jmsj/00320259}, \url{https://projecteuclid.org/journals/journal-of-the-mathematical-society-of-japan/volume-3/issue-2/Sur-les-Fonctions-Analytiques-de-Plusieurs-Variables-VIII--Lemme/10.2969/jmsj/00320259.full} (accessed 2024-03-21).
\newblock Publisher: Mathematical Society of Japan.

\bibitem{parrilo_chapter_2012}
{\sc P.~A. Parrilo}, {\em Chapter 3: {Polynomial} {Optimization}, {Sums} of {Squares}, and {Applications}}, in Semidefinite {Optimization} and {Convex} {Algebraic} {Geometry}, {MOS}-{SIAM} {Series} on {Optimization}, Society for Industrial and Applied Mathematics, Dec. 2012, pp.~47--157, \url{https://doi.org/10.1137/1.9781611972290.ch3}, \url{https://epubs.siam.org/doi/abs/10.1137/1.9781611972290.ch3} (accessed 2024-01-26).

\bibitem{shechtman_phase_2015}
{\sc Y.~Shechtman, Y.~C. Eldar, O.~Cohen, H.~N. Chapman, J.~Miao, and M.~Segev}, {\em Phase {Retrieval} with {Application} to {Optical} {Imaging}: {A} contemporary overview}, IEEE Signal Processing Magazine, 32 (2015), pp.~87--109, \url{https://doi.org/10.1109/MSP.2014.2352673}.

\bibitem{trefethen_is_2008}
{\sc L.~N. Trefethen}, {\em Is {Gauss} {Quadrature} {Better} than {Clenshaw}–{Curtis}?}, SIAM Review, 50 (2008), pp.~67--87, \url{https://doi.org/10.1137/060659831}, \url{https://epubs.siam.org/doi/abs/10.1137/060659831} (accessed 2024-01-26).

\bibitem{venugopalkrishna_fredholm_1972}
{\sc U.~Venugopalkrishna}, {\em Fredholm operators associated with strongly pseudoconvex domains in \textit{{C}}n}, Journal of Functional Analysis, 9 (1972), pp.~349--373, \url{https://doi.org/10.1016/0022-1236(72)90007-9}, \url{https://www.sciencedirect.com/science/article/pii/0022123672900079} (accessed 2024-05-30).

\bibitem{wang_solving_2018}
{\sc G.~Wang, G.~B. Giannakis, and Y.~C. Eldar}, {\em Solving {Systems} of {Random} {Quadratic} {Equations} via {Truncated} {Amplitude} {Flow}}, IEEE Transactions on Information Theory, 64 (2018), pp.~773--794, \url{https://doi.org/10.1109/TIT.2017.2756858}, \url{https://ieeexplore.ieee.org/abstract/document/8049465} (accessed 2024-04-16).
\newblock Conference Name: IEEE Transactions on Information Theory.

\bibitem{yonel_generalization_2019}
{\sc B.~Yonel and B.~Yazici}, {\em A {Generalization} of {Wirtinger} {Flow} for {Exact} {Interferometric} {Inversion}}, SIAM Journal on Imaging Sciences, 12 (2019), pp.~2119--2164, \url{https://doi.org/10.1137/19M1238599}, \url{https://epubs.siam.org/doi/abs/10.1137/19M1238599} (accessed 2024-01-26).

\end{thebibliography}

\end{document}


\maketitle

\section{Numerical Considerations}
\subsection{Cost Function, Gradient, and Hessian}
We consider the sums of squares cost function presented in the main text, along with a regularizer that fixes the global phase of the global minima. exists, 
\begin{equation}
    f(\textbf{x};\textbf{w}) = \frac{1}{2} \norm{\frac{|\mathcal{F}\{\textbf{x}\}|^2}{\sqrt{\textbf{y}}} - \sqrt{\textbf{y}}}_2^2 + \frac{\Im{\textbf{x}_\textbf{w}}}{2}^2
\end{equation} \label{eq:cost-function}
We use Wirtinger notation \cite{} for the Hessian and gradient, rather than writing out the gradient and Hessian with respect to the real and imaginary parts of $\textbf{x}$. The two approaches are equivalent; Wirtinger notation just simplifies the algebra. In the following we use $\textbf{e}_\textbf{w}$ to represent the $\textbf{w}$ - Euclidean basis vector. 
The two Wirtinger gradients are given by:
\begin{equation}
    \nabla_{\overline{\textbf{x}}} f(\textbf{x}; \textbf{w}) = \mathcal{F}^{\dagger}\left [\left (\frac{|\mathcal{F}\{\textbf{x}\}|^2}{\textbf{y}} - \textbf{1} \right) \odot \mathcal{F}\{\textbf{x}\} \right ] + \Im{\textbf{x}_\textbf{w}} \textbf{e}_\textbf{w} 
\end{equation}
\begin{equation}
    \nabla_{\textbf{x}} f(\textbf{x}; \textbf{w}) = \overline{\nabla_{\overline{\textbf{x}}} f(\textbf{x}; \textbf{w})}
\end{equation}

It is easy to see that for a noiseless scenario, $|\mathcal{F}\{\textbf{x}\}|^2 = \textbf{y}$, the gradient becomes zero, as expected for the minimum. The full Wirtinger Hessian is Hermitian and is given by:

\begin{equation} \label{eq:WirtingerHessian}
    \mathcal{H} = \begin{pmatrix}
        \mathcal{H}_{\textbf{x} \textbf{x}} & \mathcal{H}_{\overline{\textbf{x}} \textbf{x}} \\ \mathcal{H}_{\textbf{x} \overline{\textbf{x}} } &\mathcal{H}_{\overline{\textbf{x}} \overline{\textbf{x}} } 
    \end{pmatrix}
\end{equation}
The components of the Hessian, defined by their actions on a test vector $\textbf{u}$, are given by:
\begin{equation}
    \mathcal{H}_{\textbf{x} \textbf{x}} [ \textbf{u}]  =  \mathcal{F}^{\dagger} \left [\left (2\frac{|\mathcal{F}\{\textbf{x}\}|^2}{\textbf{y}} - \textbf{1} \right) \odot \mathcal{F}\{\textbf{u}\}   \right ] + \textbf{e}_\textbf{w} \textbf{e}_\textbf{w}^\dagger
\end{equation}
\begin{equation}
    \mathcal{H}_{\overline{\textbf{x}} \textbf{x}} [ \textbf{u}]  =  \mathcal{F}^{\dagger} \left [\frac{\mathcal{F}\{\textbf{x}\}^2}{\textbf{y}}\odot \mathcal{F}^{\dagger}\{\textbf{u}\}   \right ] -  \textbf{e}_\textbf{w} \textbf{e}_\textbf{w}^\dagger
\end{equation}
\begin{equation}
    \mathcal{H}_{\overline{\textbf{x}} \overline{\textbf{x}} } = \overline{\mathcal{H}_{\textbf{x} \textbf{x}}} 
\end{equation}
\begin{equation}
    \mathcal{H}_{\textbf{x} \overline{\textbf{x}} } =  \mathcal{H}_{\overline{\textbf{x}} \textbf{x}}^\dagger 
\end{equation}

The Newton system of equations in Wirtinger notation is:
\begin{equation}
    \mathcal{H} \begin{pmatrix}
        \Delta \textbf{x} \\ \Delta \overline{\textbf{x}}
    \end{pmatrix} = - \begin{pmatrix}
        \nabla_{\overline{\textbf{x}}} f \\ \nabla_{\textbf{x}} f
    \end{pmatrix}
\end{equation}
In the noiseless case for $\textbf{w} = \textbf{0}$  and  $\textbf{x}_\textbf{0}$ dominating the norm of $\textbf{x}$ we have:
\begin{equation}
    \mathcal{H}_{\textbf{x} \textbf{x}}  = \mathcal{I} + \textbf{e}_\textbf{0} \textbf{e}_\textbf{0}^\dagger
\end{equation}

\begin{equation}
    \lim_{\textbf{x}_\textbf{0} \to \infty} \mathcal{H}_{\overline{\textbf{x}} \textbf{x}} = \textbf{e}_{\textbf{0}} \textbf{e}_{\textbf{0}}^\dagger -  \textbf{e}_{\textbf{0}} \textbf{e}_{\textbf{0}}^\dagger = \textbf{0}
\end{equation}
Intuitively $\mathcal{H}_{\overline{\textbf{x}} \textbf{x}}$ represents the degree of freedom offered by global phase. Our added regularizer fixes this degree of freedom which numerically corresponds to zeroing out the relevant matrix entry. 
\subsection{Strong Convexity of the Cost Function}
We prove that the cost function \ref{eq:cost-function} is strongly convex in a neighborhood around a global minimum, with the caveat that for non-zero winding number the strong convexity parameter may vary with dimension. The resulting ill-conditioning does not change the oracle complexity, only the practical implementation of the method.
\begin{theorem}
    The cost function $f(\textbf{x}; \textbf{w})$ is strongly convex at a global minimum where $|\mathcal{F}\{\textbf{x}\}|= \textbf{y}$. 
\end{theorem}
\begin{proof}
    The proof is straightforward. We show that the Hessian is strictly diagonally dominant at a solution.  We observe that at a solution $\mathcal{H}_{\textbf{x} \textbf{x}}$ is the identity matrix plus a rank one matrix. We next show that the one-norm of any given row of the off-diagonal matrix is strictly less than one. To proceed we make use of the fact that $\mathcal{H}_{\overline{\textbf{x}} \textbf{x}}$ can be embedded in a circulant matrix. Next, for a circulant matrix its eigenvalues are given by the Fourier transform of its rows. The 1-norm of the Fourier operator is $N$, so the norm of a row satisfies:
\begin{equation}
    \norm{\textbf{c}}_1 \leq \frac{1}{N} \norm{ \frac{\mathcal{F}\{\textbf{x}\}^2}{|\mathcal{F}\{\textbf{x}\}|^2}}_1 = 1
\end{equation}
Finally, neglecting the trivial case where the Circulant matrix is the identity, it must contain non-zero elements in its off-diagonal blocks. Thus, the Toeplitz truncation strictly reduces the 1-norm, yielding:
\begin{equation}
    \norm{\textbf{c}}_1 < 1
\end{equation}
Therefore, the Hessian which has 1's along the diagonal, is strictly diagonally dominant at such a global minimum. It is thus strictly positive definite, and thus the underlying cost function is strongly convex at that point.
\end{proof}

We note that the Hessian being strictly diagonally dominant means that a wide variety of iterative methods can be successfully applied to the Wirtinger Newton system. 
\subsection{Preconditioning, and Conjugate Gradients}
\begin{figure}[H]
    \centering
    \includegraphics[scale = .75]{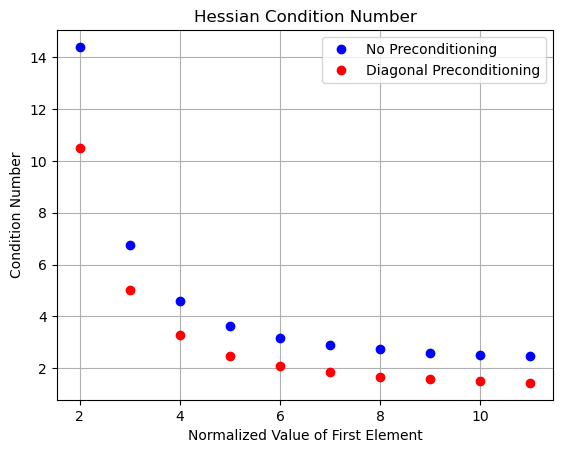}
    \caption{Condition Number vs normalized First Entry}
    \label{fig:condition_number}
\end{figure}

Figure \ref{fig:condition_number} shows the condition number of the Wirtinger Hessian with and without diagonal preconditioning, for the case that $\textbf{w} = \textbf{0}$ as we increase the magnitude of the first element of $\textbf{x}$. 

The Hessian becomes extremely well-conditioned in the asymptotic case, making the conjugate gradient (CG) method ideal for solving the Newton system.

\section{Connections to Algebraic Topology}
For those familiar with Algebraic topology, we explain how our method ties in with several classical results in the field. Our method is fundamentally an application of the Atiyah–Singer index theorem \cite{korevaar_several_nodate}. The multi-winding number we refer to in the main text corresponds to the analytic index of the Fredholm  operator corresponding to the Dirichlet problem that solves the relevant multiplicative Cousin problem \cite{korevaar_several_nodate}. This provides insight into why we need iterative refinement to improve the solution: for non-zero analytic index the Fredholm operator possesses a non-vanishing kernel, and so the Dirichlet problem is ill-posed, meaning direct analytic methods cannot be applied. Viewed from this perspective, our method can be understood as a parametrix method for the Complex Monge-Ampere equation \cite{korevaar_several_nodate} (the dirichlet problem defined by the multiplicative cousin problem).

\bibliographystyle{siamplain}
\bibliography{fastPhase_new}